\documentclass[prb, 
superscriptaddress,showpacs,amsmath,amssymb]{revtex4}
\usepackage{amsfonts}
\usepackage{bm}
\usepackage{verbatim}
\usepackage{graphicx}

\graphicspath{{pics/}}

\begin{document}

\title{Graphite on graphite}

\author{G.E.~Volovik}
\affiliation{Low Temperature Laboratory, Aalto University,  P.O. Box 15100, FI-00076 Aalto, Finland}
\affiliation{Landau Institute for Theoretical Physics RAS, Kosygina 2, 119334 Moscow, Russia}
\affiliation{P.~N. Lebedev Physical Institute of the RAS, Moscow 119991, Russia}

\author{V.M. Pudalov}
\affiliation{P.~N. Lebedev Physical Institute of the RAS, Moscow 119991, Russia}
\affiliation{National Research University Higher School of Economics, Moscow 101000, Russia}

\date{\today}

\begin{abstract}
 We propose potential geometry for fabrication  of the  graphite sheets with atomically smooth edges.
For such sheets with  Bernal stacking, the electron-electron interaction and topology should cause sufficiently high density of states resulting
in the high temperature  of
either  spin ordering or superconducting  pairing. \cite{Pisma}
\end{abstract}

\maketitle

The perspective direction in increasing the
critical temperature of superconductivity is the search for and synthesis of the electronic systems with flat band in the
energy spectrum. The dispersionless energy spectrum has a singular density of states, which provides the transition
temperature being proportional to the coupling constant instead of the exponential suppression.\cite{KhodelShaginyan1990}
There are different potential sources of flat band.

One of them is formation of the so-called Khodel-Shaginyan fermion condensate -- the flat band  caused by electron-electron interaction in metals \cite{KhodelShaginyan1990,Volovik1991,Nozieres1992}. This is the manifestation of the general phenomenon of the energy level merging due to electron-electron interaction. This effect has been recently
suggested \cite{Dolgopolov2014} to be responsible for merging  of the discrete energy levels in two-dimensional electron system in quantizing magnetic fields. Another case of merging was found on the verge of the metal-insulator transition  or quantized Hall effect to insulator transition,
where energy levels for difference subbands merge instead of repulse \cite{pudalov-Hall_JETPL_1993,krav_PRL_1955,pudalov-termination_SS_1994}.

 According to Ref.~\cite{Yudin2014} the favourable condition for the formation of such flat band is when the van Hove singularity comes close to the Fermi surface \cite{KhodelShaginyan1990} (see also Ref.~\cite{Volovik1994} for the simple Landau type model of the formation of such flat band). It is also possible that this effect is responsible for the occurrence of superconductivity with high $T$ observed in the pressurized sulfur hydride \cite{Drozdov2014,Drozdov2015}.
There are some theoretical evidences \cite{Pickett2015,Bianconi2015}  that the high-$T_c$ superconductivity takes place at such pressure, when the system is near the Lifshitz transition, i.e. when the Fermi surface is close to van Hove singularity and the bands flatten.
That is why it is not excluded that  the Khodel-Shaginyan flat band is formed  in sulfur hydride at pressure 180-200 GPa giving rise to high-T$_c$ superconductivity.

The flat band can be also formed in
such semimetals which contain the Dirac lines. Dirac nodal lines are the topologically protected lines in the energy spectrum, which have zero energy \cite{Ryu2002,HeikkilaVolovik2011,HeikkilaKopninVolovik2011,SchnyderRyu2011,HeikkilaVolovik2016}.
The flat band appears on the surface of such material due to the topological phenomenon of the bulk-edge correspondence.
The boundary of the surface flat band is determined by the projection of the Dirac line in the bulk to the plane of the surface.
By now, there are many evidences of the existence of Dirac lines in the
electronic spectrum of semimetals \cite{Weng2014,Kane2015,Belopolski2015,Cha2015,Bian2015,Zhao2015,Hirayama2016,Huang2016,LiuBalents2016}.

The Dirac nodal lines and the corresponding flat bands exist also in the pair-correlated systems: in cuprate superconductors and in the spectrum of fermionic excitations in the recently discovered polar phase of superfluid $^3$He \cite{Dmitriev2015}. They lead  to the singularity in the density of states of the fermions bound to the vortex.\cite{Volovik1993,KopninVolovik1996,Volovik2016}.

Graphite is one of the realizations of the approximate nodal line semimetal (for the review, see \cite{HeikkilaVolovik2016}).
In the model, where some small hopping elements are neglected, the Bernal graphite has two vertical Dirac lines running between the {\bf H} points in the 3D graphite band structure.  According to the
bulk-edge correspondence, the Dirac line in bulk along the normal to the graphite planes produces  the zero energy states on the lateral boundaries of Bernal graphite, which form the surface flat band. This 2D flat band can be considered as extension of the 1D flat band states in  graphene with zigzag edges \cite{HeikkilaVolovik2016}. Note that in the similar model of the rhombohedral graphite the Dirac line forms a spiral, whose projection on the horizontal boundary marks the boundary of the flat band.
In the Bernal graphite model the flat band exists only on the lateral boundary.

In the real graphite the flat band is distorted due
to two reasons.
The first is due to the higher-order hopping elements, which are neglected in the model. These elements break the symmetry, which supports the topological stability of the Dirac line. In real Bernal and rhombohedral graphite the Dirac line transforms to the electron and hole pockets \cite{McClure1957}.  The topology of the Dirac lines, however, is not fully destroyed: the electron and hole pockets form the continuous chain. Nevertheless, the extension of the Dirac line to the chain of Fermi surfaces distorts the surface flat band. Whether such distortion of the flat band considerably reduces the high transition temperature expected for the flat band materials is to be investigated both numerically and experimentally. The flattening of the spectrum for the five layers of graphene
with the rhombohedral ABCAB stacking has been observed experimentally \cite{Pierucci2015} by
 epitaxial  growing  on  3C-SiC(111)  on  2$^�$off�axis  6H-SiC(0001).  Note that the there can be the combined effect of topology and electron-electron interaction for the stabilization of the flat band, i.e. due to the interaction a part of the approximate flat band may become exact.

However, there is another important source of distortion of the edge flat band in Bernal graphite: the lateral walls of graphite are always rough. This is not important for the rhombohedral graphite, where the (approximate) Dirac lines are not vertical and thus have nonzero projection on the smooth horizontal boundary. That is why we need such configuration of the Bernal graphite, where the lateral walls can be made smooth.

 In order to fabricate smooth lateral walls, one obvious way is to cut the graphite edge with a nanometer precision
by using scanning tunneling lithography \cite{tapas_Nat.Nano_2008,biror_2010}; in this way  graphene nanoribbons with  smooth edges were obtained
in Ref.~\cite{Wang PRL 2008}.
The graphite flakes may be shaped using  Ar- or  He-ion beam etching.
 
Without relying on lithographic techniques,   fabrication of nanoribbons with atomically smooth edges has also been reported via the so-called ``bottom-up'' processes
-- epitaxial growth
on templated  nanofacets \cite{sprinkle_2010} and by chemical
vapor deposition  on surface features  \cite{ago_2012,hayashi_2012,ago_2013}. Recently,  graphene nanoribbons with atomically flat edges have been grown  on a semiconductor Ge-wafer \cite{jacobberger_NatCom_2016}.

Another way to grow graphene/graphite sheets orthogonal to the atomically flat graphite  substrate is shown schematically in  Fig.~1.
In order to facilitate such
growth, an atomically thin and narrow strip of a catalyst (Pt, Ni or TiC) may be deposited at the flat surface of a crystalline substrate (graphite, (001)Ge, h-BN, etc) in the proper direction. Alternatively, growth in the out-of-plane direction may be initiated by surface features such as trenches, steps or ridges.
Then, in the CVD reactor the graphite sheets may  start growing perpendicular to the substrate plane and directed along the strip.

The growth conditions in the proposed geometry, of cause, require more thorough consideration.
We note, however, that it is similar to the ``carbon nanowalls'' which are graphite nanostructures with edges
comprised of stacked planar graphene sheets
standing vertically on a substrate. The sheets form a wall structure with thicknesses
in the range of a few nanometers to a few tens of nanometers \cite{hiramatsu_book}.

\begin{figure}[top]
\centerline{\includegraphics[width=0.5\linewidth]{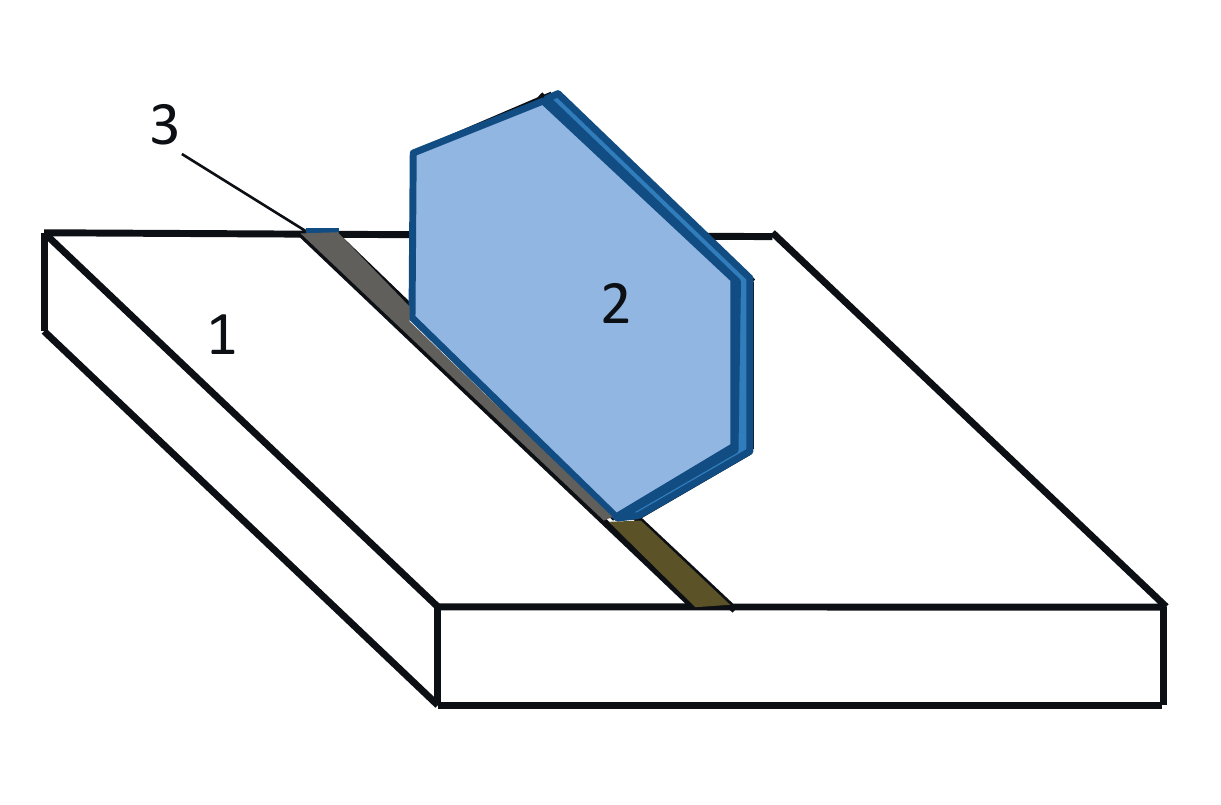}}
\label{GoG} 
  \caption{Schematic view of the graphite-on-graphite design. {\em 1} -- graphite (BN) substrate, {\em 2} - graphite (or a multilayered graphene) sheet, or a graphite wall,
{\em 3} --  a strip of the catalist.
 }
\end{figure}

In the case of success,  the interface between the vertical graphite and the horizontal graphite layers may be made smooth.
With the appropriate mutual orientations of the interface and crystal axes, the flat interface contains the (approximate) flat band, and may have sufficiently large density of states to produce the high temperature of transition to either magnetic or superconducting state. Graphene itself also may serve as substrate.

At the moment there are many evidences of enhanced superconducting transition temperature, which are related to monolayers \cite{FeSe}, interfaces \cite{Esquinazi2014,Esquinazi2013a,Esquinazi2013b,Esquinazi2013c,TangFu2014,Bozovic},
alkanes in contact with a graphite surface \cite{Kawashima2013},
 polymer composites with embedded graphene flakes \cite{Ionov2015,Ionov2016}, etc.
The interface itself may provide the factor which together with the electron-electron interaction enhances the
effect of the flat band singularity in the electronic density of states leading to unexpected effects.

VP benefited from discussions with K.~N.~Eltsov and E.~D.~ Obraztsova.
The authors acknowledge support by RSCF (\# 16-42-01100).


\end{document}